\documentclass[%
 reprint,
 amsmath,amssymb,
 aps,
]{revtex4-2}

\usepackage{graphicx}
\usepackage{dcolumn}
\usepackage{bm}

\usepackage{url}

\usepackage{algorithmic}
\usepackage{algorithm}
\usepackage{booktabs}
\usepackage{multirow}

\begin{document}

\preprint{APS/123-QED}

\title{Machine-Learning-Based Exchange–Correlation Functional 
\\with Physical Asymptotic Constraints}

\author{Ryo Nagai$^{1,2}$}
\email{r-nag@issp.u-tokyo.ac.jp}
\author{Ryosuke Akashi$^{2}$}
\author{Osamu Sugino$^{1,2}$}
\affiliation{$^1$The Institute for Solid State Physics, The University of Tokyo, Kashiwa, Chiba 277-8581, Japan}
\affiliation{$^2$Department of Physics, The University of Tokyo, Hongo, Bunkyo-ku, Tokyo 113-0033, Japan}

\date{\today}

\begin{abstract}
Density functional theory is the standard theory for computing the electronic structure of materials, which is based on a functional that maps the electron density to the energy. However, a rigorous form of the functional is not known and has been heuristically constructed by interpolating asymptotic constraints known for extreme situations, such as isolated atoms and uniform electron gas. Recent studies have demonstrated that the functional can be effectively approximated using machine learning (ML) approaches. However, most ML models do not satisfy asymptotic constraints. In this study, by applying a novel ML model architecture, we demonstrate a neural network-based exchange-correlation functional satisfying  physical asymptotic constraints. Calculations
reveal that the trained functional is applicable to various materials with an accuracy higher than that of existing
functionals, even for materials whose electronic properties are  different from the properties of materials in the training dataset. Our proposed approach thus improves the accuracy and  generalization performance of the ML-based functional by combining the advantages of ML and
analytical modeling.
\end{abstract}

\maketitle


\section{\label{sec:intro}Introduction}
Density functional theory (DFT) \cite{HK} is a method for electronic structure. It is useful for elucidating the physical properties of various materials and plays an important role in industrial applications, such as drug discovery and semiconductor development. In principle, the electronic structure can be obtained by wave function theory (WFT), which directly solves the Schrödinger equation. However, the computational cost increases exponentially with respect to the numberof electrons, which prohibits computation for materials with tens of atoms or more. Alternatively, the electronic structure can be obtained by solving the Kohn–Sham (KS) equation in DFT with a cubic computational cost \cite{KS}. This makes it possible to compute the electronic structure of materials with up to millions of atoms. As a result, DFT has become for calculating the electronic structure.

However, at the cost of the reduced computational complexity, the KS equation includes a term whose rigorous form is unknown. This term is called the exchange-correlation (XC) functional, which relates the electron density to the many-body interaction of electrons. This term has been heuristically approximated in various ways. The most common form of the XC functional is a combination of several asymptotic limits formulated with physical theories, such as the uniform electron gas limit \cite{Slater}. It has been argued that satisfying as many physical constraints as possible is important to increase the performance of the approximation \cite{DFT_is_straying}. However, the combination of physical constraints is often performed with analytically tractable models, which may be insufficient to represent the intrinsic complexity of the XC functional. 

In this context, another approach has emerged, namely, the application of machine learning (ML). By training a flexible ML model with high-quality material data, one can construct XC functionals containing more complex electronic interactions than by the analytical modeling. This approach has been demonstrated to yield successful results in various systems. \cite{Snyder, Burke-bypassing, NNVHxc}
Brockherde {\it et al.} proposed a ML mapping to obtain electronic structures with lower computational cost than by the KS equation \cite{Burke-bypassing}.  This method was demonstrated to be effective in accelerating materials simulations; however, it can only be applied to materials composed of the same chemical elements as those used in training. Nagai {\it et al.} applied ML to improve the accuracy of the XC functional in the KS equation \cite{NNVHxc, nagai2020completing}. Their ML models are designed to be applicable to materials containing any chemical element; however, its performance becomes
unreliable for materials that differ significantly from the materials in the training dataset. In some cases, the convergence of the calculation becomes unstable. 

These problems can be solved by including all archetypal materials with 
all chemical elements in the training dataset; however, this is impractical because of the difficulty of obtaining such a large amount of data. Because it is difficult to obtain electronic states with sufficient accuracy via experiments, training data are mainly generated by calculation. 
To exceed the accuracy of conventional DFT, the data should be generated using WFT because its accuracy must be higher than that of DFT with the approximated XC functional. However, the computational complexity of modern WFT is $\mathcal{O}(N^7)$ or higher, where $N$ denotes the system size. Thus, the materials available for training are limited to small molecules. The greatest challenge in the ML-based modeling is to achieve high generalization performance with limited data.

As a solution to this problem, a method of imposing physical constraints on  ML functionals has been proposed.
Hollingsworth {\it et al.}  imposed a physical constraint on the ML-based non-interacting kinetic energy functional in a 1D model system  \cite{kieron_exact_condition}. Without the physical constraint, the constructed kinetic energy functional could take any possible electron density as its input and predict the corresponding energy even if it did not resemble the training data. Alternatively, by imposing a coordinate-scaling condition on the density, Hollingsworth {\it et al.} limited the target of ML to density with a specific spatial width. By transforming the output of the ML model with the scaling condition, they obtained the kinetic energy applicable to electron densities with various system widths. Their results demonstrated that ML models with the coordinate-scaling
condition learned the system with higher accuracy than ML models without constraints. This method reduced the burden on ML by compressing the dimensionality of the target system. Therefore, by introducing physical constraints appropriately, the accuracy and generalization performance of ML models can be improved.

As an extension of this approach, we propose a method to construct a ML-based XC functional for real three-dimensional materials satisfying physical asymptotic  constraints.
Several asymptotic dependencies of the XC functional in 3D systems have been derived analytically, which various approximate functionals have been developed to satisfy. 
The importance of physical constraints in improving transferability has been argued \cite{DFT_is_straying}. 
We believe that even in the construction of a ML XC functional, applying appropriate physical constraints regularize the behavior of the functional when there is insufficient training.


In this paper, we propose a novel architecture to allow a ML model to satisfy the asymptotic constraints. 
We apply this approach to XC functional construction based on a neural network (NN), and present training and benchmarking of the NN-based XC functional that satisfies as many physical asymptotic constraints as possible. 

\renewcommand{\arraystretch}{1.3}
\begin{table*}[t]
\caption{Applied physical constraints. Numbers starting with X denote the constraints applied to the exchange part, while those starting with C denote the constraints applied to the correlation part. Items in column ${\bf x_0}$ correspond to the input of meta-generalized
gradient approximation models: $(\bar{s}, \bar{\tau})$ for X, and $(\bar{n}_s, \bar{\zeta}, \bar{s}, \bar{\tau})$ for C (see Eqs. \ref{eq:fx} and \ref{eq:fc}). Column $f_0$ presents the asymptotic limits that $F_{\rm X, C}$ should converge to.}
 \label{table:applied_conditons}
\begin{ruledtabular}
\centering
  \begin{tabular}{cccc}
    & Physical constraint & ${\bf x_0}$ & $f_0$ \\
   \hline
   X1 &Correct uniform coordinate density-scaling behavior &\multirow{2}{*}{-}&\multirow{2}{*}{(applied by Eq. \ref{eq:spinrelation} and \ref{eq:fx})}\\
   X2 & Exact spin scaling relation &  & 
   \\
   X3 & Uniform electron gas limit &  $(0, 0)$ & 1
   \\
   X4 &$F_{\rm X}$ vanishes as $s^{-1/2}$ at $s\rightarrow\infty$  & $(1, \bar{\tau})$ & 1
   \\
   X5 & Negativity of $\varepsilon_{\rm X}$&-&(applied by Eq. \ref{eq:final_processing}) 
   \\
   C1 & Uniform electron gas limit & ($\bar{n}_s$, $\bar{\zeta}$, 0, 0) &1
   \\
   C2 & Uniform density scaling to low-density limit &\multirow{2}{*}{
   (0, $\bar{\zeta}$, $\bar{s}$, $\bar{\tau}$)
   } 
   & \multirow{2}{*}{
   $F_{\rm C}^{\rm NN}(\bar{n}_s, 0, \bar{s}, \bar{\tau})-F_{\rm C}^{\rm NN}(0, 0, \bar{s}, \bar{\tau}) + 1$
   }
   \\
   C3 & Weak dependence on $\zeta$ in low-density region
   & &
   \\
   C4 & Uniform density scaling to high-density limit & (1, $\bar{\zeta}$, $\bar{s}$, $\bar{\tau}$) & 1 
   \\
   C5 & Nonpositivity of $\varepsilon_{\rm C}$ & -&(applied by Eq. \ref{eq:final_processing})  \\
  \end{tabular}
 \end{ruledtabular}
\end{table*}

\section{Methodology}
\subsection{Imposing Asymptotic Constraints on ML Models}
Suppose that we want  a ML model $f$ to satisfy an asymptotic constraint $f({\bf x})\rightarrow f_0$ at ${\bf x}\rightarrow {\bf x}_0$. One approach is to have the training data include values at that asymptotic limit. However, ML models trained in this way do not always take the exact asymptotic value, unless the training data is strictly fitted. Alternatively, we propose making ML model satisfy the constraint in an analytical way. Instead of training the output of the ML function $f$ directly, we train the following processed form: 
\begin{equation}
    \hat{\theta}_{\{{\bf x}_0,f_0\}}[f]({\bf x})=f({\bf x})-f({\bf x}_0)+f_0.
    \label{eq:single}
\end{equation}
Here, this operation is represented by $\hat{\theta}_{\{{\bf x}_0,f_0\}}$. The same ML model is substituted for the first and second terms on the right-hand side.
This operator forces the model to satisfy the constraint $f(x_0) = f_0$ strictly.

Furthermore, suppose that we have $N_c$ constraints, where $f$ converges to $f_0^{(1)}, f_0^{(2)}, ... f_0^{(N_c)}$ at ${\bf x}_0^{(1)}, {\bf x}_0^{(2)},...,{\bf x}_0^{(N_c)}$. We propose the following formula, which, constructed as a generalization of the Lagrange interpolation,  satisfies all the constraints:
\begin{equation}
    \hat{\Theta}_{i=1,...N_c}[f]({\bf x})
    =
    \sum_{i=1}^{N_c} 
    \prod_{j\neq i}
    \frac{l_{i,j}({\bf x}-{\bf x}_j)}{l_{i,j}({\bf x}_i-{\bf x}_j)} 
    \hat{\theta}_{\{x_0^{(i)},f_0^{(i)}\}}[f]({\bf x})
    \label{eq:multi}
\end{equation}
Here, we represent this operation by $\Theta$. $l_{i,j} (i, j=1,2,...N_c)$ are function satisfying $l_{i,j}({\bf x})=0$ at $|{\bf x}|=0$. Depending on the form of $l_{i,j}$, the behavior around each asymptotic limit differs. Although further tuning can be performed,  we set $l_{i,j}(x)=\tanh(|{\bf x}|^2)$ for all $\{i,j\}$ in this work.

\subsection{KS Equation in DFT}
Here we demonstrate the effectiveness of this architecture by applying it to the problem of constructing an XC functional. Based on DFT, the electronic structure of materials can be calculated by solving the following KS equations:
\begin{equation}
\left\{-\frac{\nabla^2}{2} + V_{\rm ion}({\bf r}) + V_{\rm H}({\bf r}) + V_{\rm XC}({\bf r}) \right\} \varphi_{i}({\bf r})
\!=\!
\epsilon_{i} \varphi_{i}({\bf r}),
\label{eq:KS}
\end{equation}
\begin{equation}
V_{\rm H }({\bf r})= \int d{\bf r}' \frac{n({\bf r}')}{|{\bf r}-{\bf r}'|},
\label{eq:hartree}
\end{equation}
\begin{equation}
n({\bf r})= \sum_{i=1}^{N_{\rm occ}} |\varphi_i({\bf r})|^2.
\label{eq:density}
\end{equation}
Eq. \ref{eq:KS} is an eigenvalue equation, where the $i$-th eigenfunction and eigenvalue are represented by $\varphi_i$  and $\epsilon_i$, respectively. $N_{\rm occ}$ represents the number of occupied orbitals of the material, while $V_{\rm ion}$ is the Coulomb potential, which depends on the atomic position of each material. The forms of $V_{\rm H }$ and $V_{\rm XC}$ depend on the density $n$. The solution of this equation yields $n$; however, at the same time, $n$ is necessary for the construction of the equation itself. Therefore, the equation should be solved self-consistently; that is, the equation is solved iteratively, updating the terms using $n$ obtained in the previous step, until $n$ converges numerically.  After solving the KS equation, the total energy $E$ can be obtained as follows:
\begin{equation}
\begin{split}
    E=\sum_{i=1}^{N_{\rm occ}}\epsilon_{i}-\frac{1}{2}\int\!d{\bf r}\int\!d{\bf r}' \frac{n({\bf r})n({\bf r}')}{|{\bf r}-{\bf r}'|}
    \\
    -\int \! d{\bf r} n({\bf r}) V_{\rm XC}({\bf r}) + E_{\rm XC}[n].
\end{split}
\label{eq:Etot}
\end{equation}
$E_{\rm XC}$ and $V_{\rm XC}$ are called XC functionals, and  follow the relation below:
\begin{equation}
V_{\rm XC}({\bf r})=\frac{\delta E_{\rm XC}[n]}{\delta n({\bf r})}.
\label{eq:Vxc}
\end{equation}

The rigorous form of the XC functional as an explicit functional of $n$ is unknown. In practical use, the functional is approximated in various ways \cite{Slater, VWN, PBE, TPSS, SCAN, M06}, and the accuracy of the DFT calculation depends on the quality of this approximation.

\begin{figure*}[t]
\centering
\includegraphics[width=2.0\columnwidth]{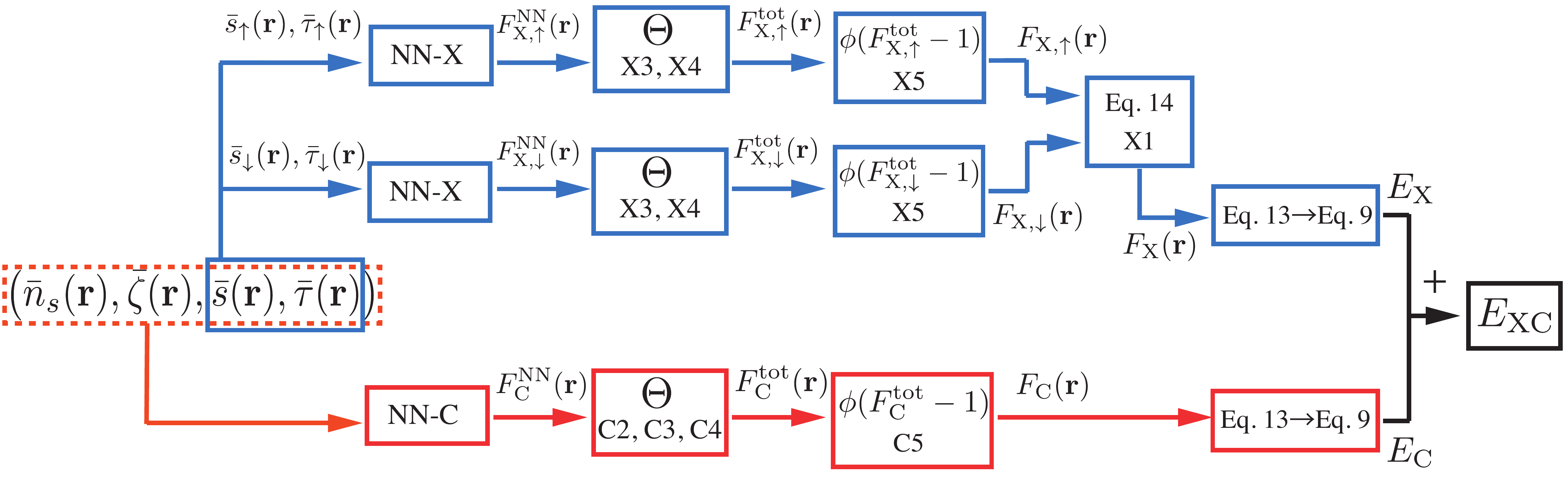} 
\caption{Schematic diagram of the complete architecture of the physically constrained neural network XC functional. $\Theta$ represents the operation to apply the asymptotic constraints presented in Eqs. \ref{eq:single} and \ref{eq:multi}, while $\phi$ represents the activation function expressed in Eq. \ref{eq:softplus}. }
\label{fig:total_architecture}
\end{figure*}

\begin{figure}[t]
\centering
\includegraphics[width=1\columnwidth]{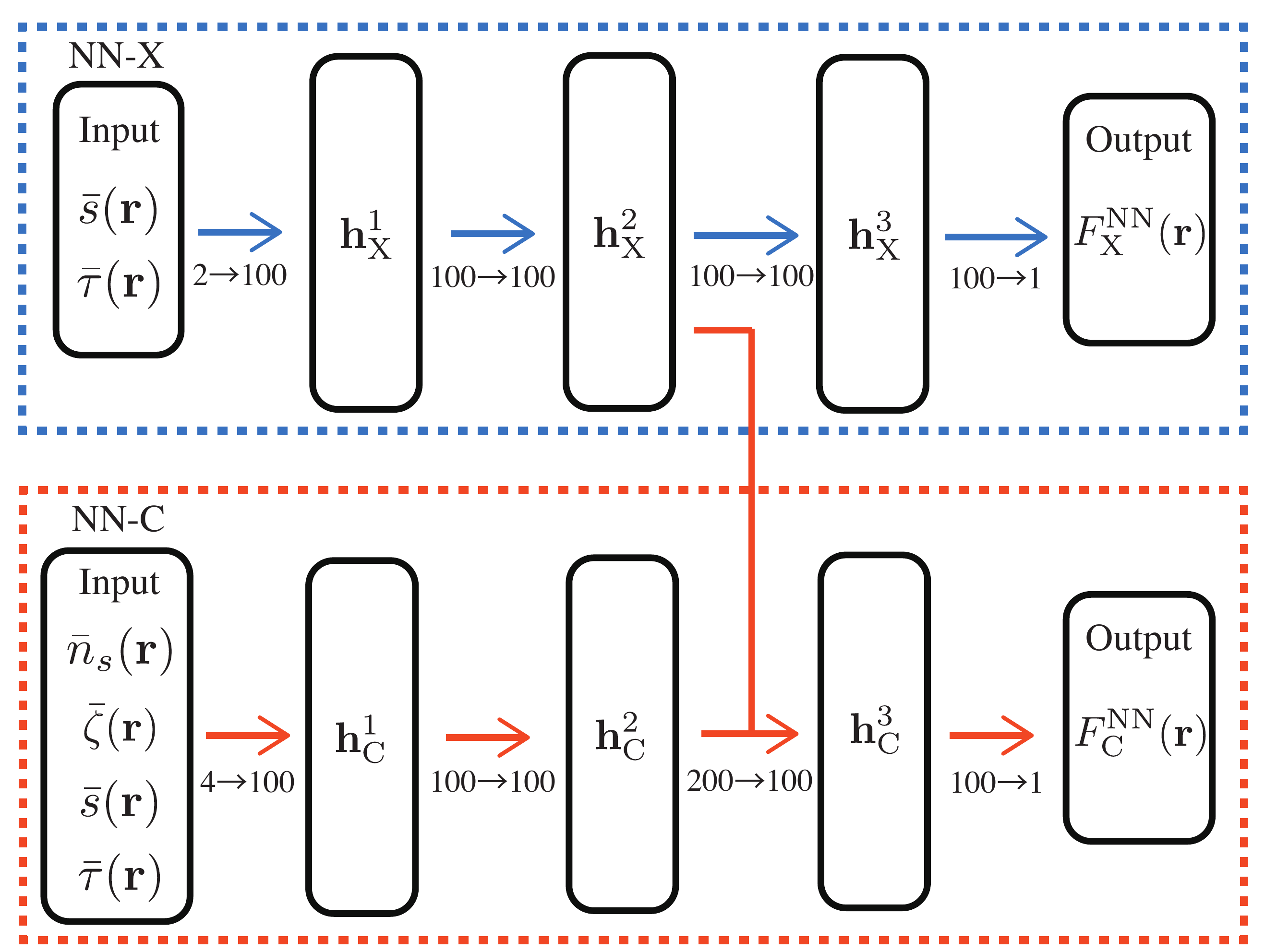} 
\caption{Schematic diagram of the neural-network architecture. The middle layers ${\bf h}_{\rm X}^2$ and ${\bf h}_{\rm C}^2$ are concatenated to compute ${\bf h}_{\rm C}^3$. The numbers between the layers represent the dimension of the matrices connecting the layers.}
\label{fig:NN_architecture}
\end{figure}

\subsection{Interpreting the XC Functional as a ML Model}
Here, we describe the method for constructing the ML-based XC functional. First, we divide the XC functional into exchange (X) and correlation (C) parts:
\begin{equation}
E_{\rm XC}[ n]= E_{\rm X}[n] +E_{\rm C}[n].
\label{eq:XandC}
\end{equation}

Instead of taking the total X and C energy as the approximation target, we approximate the XC energy density $\varepsilon_{\rm XC}$:
\begin{equation}
E_{\rm X, C}[ n]= \int d{\bf r} ~n ({\bf r})\varepsilon_{\rm X, C}[n]({\bf r}).
\label{eq:energydensity}
\end{equation}
Subscripts “X, C” indicates the equation applies to the X and C parts, respectively. By properly constructing the local form $\varepsilon_{\rm XC}$, the functinoal can be applied to systems of any size. 
Furthermore, for the convenience of calculation, most conventional studies have adopted the semilocal approximation, where $\varepsilon_{\rm X, C}[n]({\bf r})$ depends locally on the electron density around ${\bf r}$. In this study, we adopt the a meta-generalized gradient approximation (meta-GGA), where $\varepsilon_{\rm X, C}[n]({\bf r})$ refers to the following local density features \cite{TPSS}:
\begin{equation}
    \begin{aligned}
          &n_s({\bf r})&=&\quad n({\bf r})^\frac{1}{3}, \\
          &\zeta({\bf r})&=&\quad\frac{n_\uparrow({\bf r}) - n_\downarrow({\bf r})}{n({\bf r})},\\
          &s({\bf r})&=&\quad\frac{|\nabla n({\bf r})|}{n({\bf r})^{\frac{4}{3}}}, \\
          &\tau({\bf r})&=&\quad\frac{1}{2}\sum_{i=1}^{N_{\rm occ}}|\nabla\varphi_i({\bf r})|^2,
    \end{aligned}
    \label{eq:metagga}
\end{equation}
where $n_\uparrow$ and $n_\downarrow$ represent the density of electrons with each spin, and $n=n_\uparrow+n_\downarrow$.
Furthermore, in this work, each feature value is processed as follows:
\begin{equation}
    \begin{aligned}
          &\bar{n}_s({\bf r})&=&\quad\tanh (n_s({\bf r})), \\
          &\bar{\zeta}({\bf r})&=&\quad\tanh\left( \frac{1}{2}\left\{(1+\zeta({\bf r}))^\frac{4}{3}+(1-\zeta({\bf r}))^\frac{4}{3}\right\} \right),\\
          &\bar{s}({\bf r})&=&\quad\tanh({s}({\bf r})), \\
          &\bar{\tau}({\bf r})&=&\quad\tanh\left(\frac{\tau({\bf r})-\tau_{\rm unif}({\bf r})}{\tau_{\rm unif}({\bf r})}\right),
    \end{aligned}
    \label{eq:processed_metagga}
\end{equation}
where $\tau_{\rm unif}$ is the value that $\tau$ converges to the uniform electron gas limit:
\begin{equation}
    \tau_{\rm unif}({\bf r})=\frac{3}{10}(3\pi^2)^\frac{2}{3} n({\bf r})^{\frac{5}{3}}.
    \label{eq:tauunif}
\end{equation}
“$\tanh$” conversions are used for the following reasons.
First, it standardizes the features by arranging their variable range to be the same. Second, it allows us to deal numerically with asymptotic constraints at infinity. The limit $x\rightarrow\infty$ is converted to the finite limit $\tanh(x)\rightarrow1$.

Instead of training the entire XC functional, we designed a ML model as the factor from an existing analytical functional, the strongly constrained and appropriately normed (SCAN) functional \cite{SCAN}. 
\begin{equation}
    \varepsilon_{\rm X, C}[n] = \varepsilon_{\rm X, C}^{\rm SCAN}[n] F_{\rm X, C}[n]
    \label{eq:SCANcoef}
\end{equation}
SCAN satisfies the largest number of physical constraints among existing meta-GGA approximations and is one of the most popular analytical XC functionals due to its accuracy. Using a well-constructed analytical functional can help in pre-training.

According to the correct uniform coordinate density-scaling condition \cite{uniform_density_scaling}, $\varepsilon_{\rm X}$ is  independent of $\bar{n}_s$. Furthermore, by the exact spin scaling relation \cite{spin_scaling}, the spin dependency of X energy satisfies the following relation:
\begin{equation}
    E_{\rm X}[n_\uparrow, n_\downarrow]= \frac{1}{2}(E_{\rm X}[ 2n_\uparrow]+E_{\rm X}[ 2n_\downarrow]).
    \label{eq:spinrelation}
\end{equation}
For these constraints, we simply need to construct $F_{\rm X}$ as a function that depends only on $\bar{s}$ and $\bar{\tau}$:
\begin{equation}
    F_{\rm X}[ n]({\bf r})= F_{\rm X}(\bar{s}({\bf r}), \bar{\tau}({\bf r})).
    \label{eq:fx}
\end{equation}
In contrast, the C part depends on all four features:
\begin{equation}
    F_{\rm C}[ n]({\bf r})= F_{\rm C}(\bar{n}_s({\bf r}), \bar{\zeta}({\bf r}), \bar{s}({\bf r}), \bar{\tau}({\bf r})).
    \label{eq:fc}
\end{equation}

In this study, we implemented $F_{\rm X, C}$ using NNs. Here, we note that the differentiation in Eq. \ref{eq:Vxc} can be implemented using automatic differentiation NN libraries. We used JAX \cite{jax2018github} to implement the NN.

\subsection{Imposing Physical Constraints on NN-based XC functional}

Here, we demonstrate how to construct a physically constrained NN (pcNN)-based XC functional. Figures \ref{fig:total_architecture} and  \ref{fig:NN_architecture} present the schematic diagrams of the architecture. For both the X and C parts, we used a fully connected NN with three hidden layers, with each layer containing 100 nodes. We referred to the optimal NN size reported by Nagai et al. \cite{nagai2020completing}. The middle layer of the NN of the X part was combined with that of the C part to efficiently share possible common characteristics between them.

For the constructed NNs $F_{\rm X}^{\rm NN}$ and $F_{\rm C}^{\rm NN}$, we imposed the asymptotic constraints using the operation $\Theta$.
The physical constraints applied in this study are listed in Table \ref{table:applied_conditons}.
Constraints X3 \cite{Slater} and X4 \cite{perdew2014gedanken} were applied for the X part, while constraints C1 \cite{VWN}, C2 \cite{SCAN}, C3 \cite{TPSS, perdew1992accurate}, and C4 \cite{levy1991density} were applied for the C part. Since the XC energy density of SCAN in Eq. \ref{eq:SCANcoef} satisfies constraints X3, X4, C1, C2, and C4, the total $\varepsilon_{\rm XC}$ can satisfy them by converging to $f_0 =1$. In $f_0$ of the constraint C3, the output of the NN with $\bar{\zeta}=0$ ($F_{\rm C}^{\rm NN}(\bar{n}_s, 0, \bar{s}, \bar{\tau})$) is used to suppress the $\zeta$-dependency. By cancelling the first and second terms, $f_0$ converges to 1 at the low-density limit, and thus satisfies  constraint C2.

\subsection{Loss Functions}
The loss function consists of the atomization energiy (AE) and electron densities of ${\rm H_2O}$, ${\rm NH_3}$, and ${\rm CH_2}$ molecules, and the ionization potential (IP) of ${\rm H_2O}$:
\begin{align}
    \Delta=
    &c_1(\Delta_{\rm AE}^{\rm H_2O}+\Delta_{\rm AE}^{\rm NH_3}+\Delta_{\rm AE}^{\rm CH_2})/{\rm AE}_0\nonumber\\
    &+c_2\Delta_{\rm IP}^{\rm H_2O}/{\rm IP}_0
    +c_3(\Delta_n^{\rm H_2O}+\Delta_n^{\rm NH_3}+\Delta_n^{\rm CH_2}),\nonumber\\
\label{eq:lossfunc}
\end{align}
where $\Delta_{\rm AE}$, $\Delta_{\rm IP}$, and $\Delta_n$ represent errors between the calculated and reference values of the AE, IP, and density, respectively. Both ${\rm AE}_0$ and ${\rm IP}_0$ were set to 1 hartree. In this work, the coefficients are set to $c_1=1, c_2=1$, and $c_3=10$ to make the average magnitudes of all the terms similar.
The reference atomization energies, ionization energies, and electron densities were obtained by the coupled-cluster singles and doubles plus perturbative triples (CCSD(T)) method \cite{CCSD}. 
The inclusion of the electron density, a continuous quantity in 3D space, provided a large amount of information for training the weights of the NN. 
The error of the electron density was defined as follows:
\begin{align}
    \Delta_n=\frac{1}{N_e}\sqrt{\int d{\bf r}~\left( n^{\rm DFT}({\bf r})- n^{\rm CCSD(T)}({\bf r})\right)^2},
    \label{eq:densityerror}
\end{align}
where $N_e$ represents the number of electrons of the target material.
We used PySCF \cite{PYSCF} to perform the DFT calculation to compute the energies and densities. We also used numerical integration in Eq. \ref{eq:densityerror}. The number of numerical grids was on average 30,000 points per molecule. Using this loss function, we trained the pcNN. See Appendix \ref{app:training} for the detailed training procedures \footnote{ The trained NN parameters and codes to call the XC functional are available on https://github.com/ml-electron-project/pcNN_mol (PySCF) and https://github.com/ml-electron-project/pcNN_solid (patch program for VASP source). }.

\subsection{Benchmark Settings}
We evaluated the performance of the trained pcNN by using it as an XC functional in the KS equation to investigate the accuracy of the electronic structure of various materials. As the test dataset, we referred to the AEs of 144 molecules, which were the same as those reported in the Training section in Ref. \cite{G2H0}. The DFT calculations were performed using PySCF. Additionally, because the performance for solid-state materials is also important, we conducted a test for the lattice constants of 48 solids including insulators, semiconductors, and metals, provided by \cite{VASPbench}. The calculations for solids were performed using the Vienna Ab-initio Simulation Package (VASP) \cite{vasp1, vasp2}, which is commonly used software program for calculating periodic systems. We used the pseudopotentials supplied by VASP, which are based on a projector augmented wave method and adjusted on the Perdew–Burke–Ernzerhof (PBE) functional \cite{PBE, kresse1999ultrasoft, blochl1994projector}. For the convergence of the KS cycle, the commutator direct inversion of the iterative subspace method \cite{C-DIIS} and the Pulay mixing method \cite{pulay1980convergence} with default  parameter settings were used in PySCF and VASP respectively. The list of materials and pseudopotentials used for each element are provided in the Supplementary Information.

\renewcommand{\arraystretch}{1}
\tabcolsep = 2.5pt
\begin{table}[t]
\centering
 \caption{Benchmark results for atomization energies of 144 molecules.
 (ME, mean error; MAE, mean absolute error; MRE, mean relative error; MARE, mean absolute relative error)}
 \label{table:benchmark_molecules}
\begin{ruledtabular}
  \begin{tabular}{ccccc}
   XC & \small{ME(kcal/mol)} & \small{MAE(kcal/mol)} & \small{MRE(\%)} & \small{MARE(\%)} \\
   \hline
   PBE & 16.2 & 17.3 &4.57 &5.21 \\
   SCAN &-4.5 & 6.2 &-1.02&2.16 \\
   NN-based & 1.8 & 4.8 & 1.56 & 2.24\\
   pcNN-based & -1.5 & \bf{3.8} &-0.30 &\bf{1.74} \\
  \end{tabular}
  \end{ruledtabular}
\end{table}

\tabcolsep = 3pt
\begin{table}[t]
\centering
 \caption{Benchmark results for lattice constants of 48 solids. Parentheses in the “NN-based” row indicate that, the numerical calculations for six solids did not converge; thus, only the converged calculations were used for the  statistics. } \label{table:benchmark_solids}
 \begin{ruledtabular}
  \begin{tabular}{ccccc}
   XC & \small{ME(m\AA)} & \small{MAE(m\AA)} & \small{MRE(\%)} & \small{MARE(\%)} \\
   \hline
   PBE & 33.9 & 38.1 & 0.67&0.80 \\
   SCAN &-7.5 & 22.3 &-0.26 &0.54 \\
   NN-based & (0.8) & (22.9)&(-0.08) &(0.53) \\
   pcNN-based & 0.5 & \bf{19.5} &-0.06 &\bf{0.46} \\
  \end{tabular}
  \end{ruledtabular}
\end{table}

\begin{figure*}[t]
\centering
\includegraphics[width=2\columnwidth]{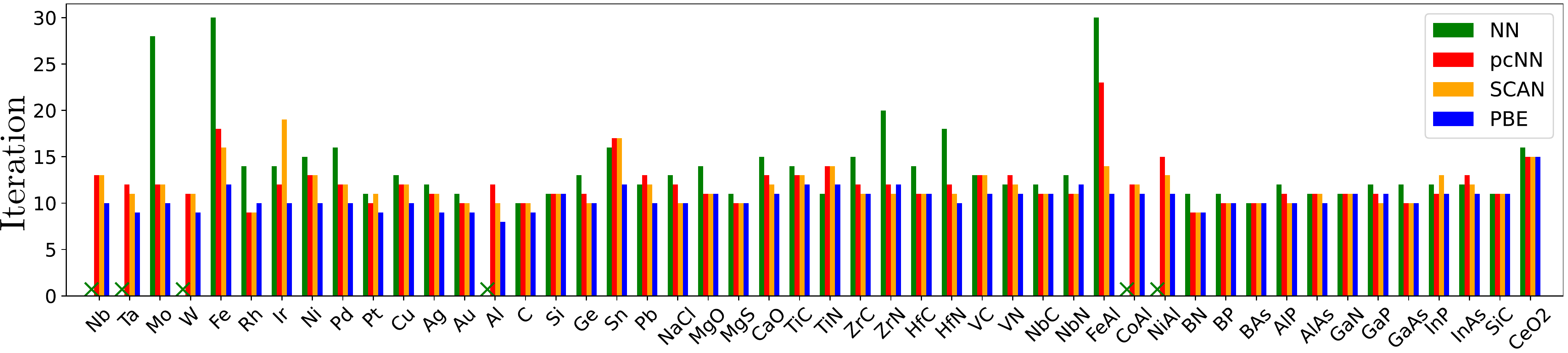} 
\caption{Comparison of the convergence of each functional for all benchmark solids. The green x symbols indicate that the NN-based functional did not converge for the corresponding material.}
\label{fig:convergence_all}
\end{figure*}

\begin{figure*}[t]
\centering
\includegraphics[width=2\columnwidth]{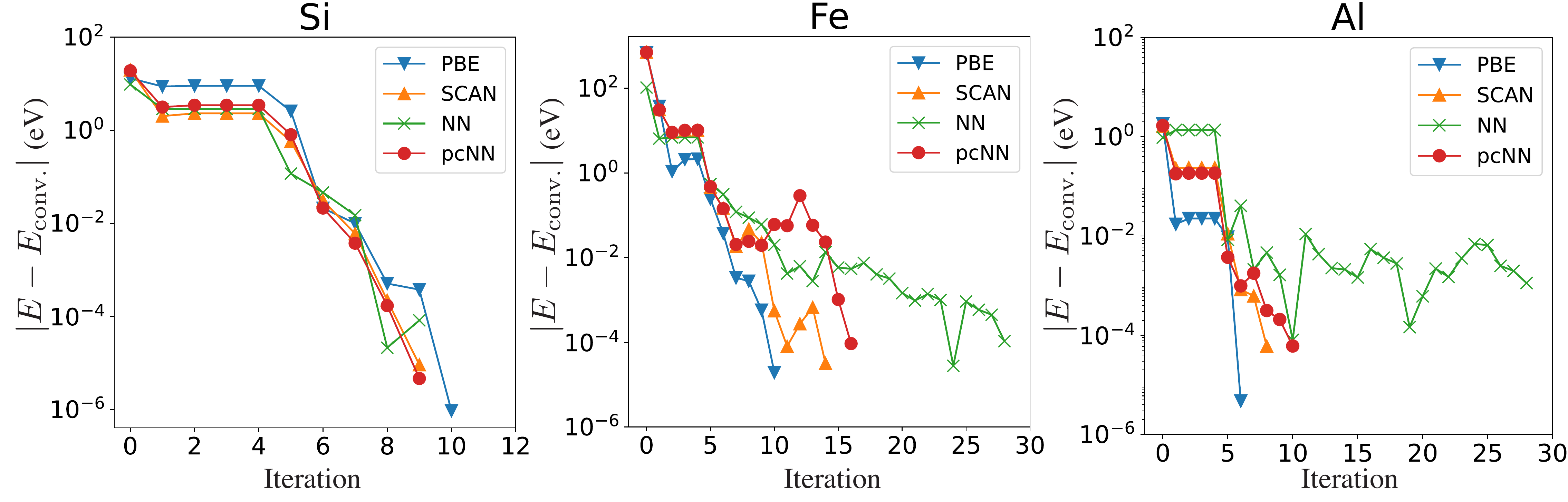}
\caption{Comparison of the total-energy convergence of each functional for Si, Al, and Fe. The vertical axis displays the difference between the energy of the current step and the energy of the final iteration $E_{\rm conv.}$. The energy value at the 30th iteration was used for $E_{\rm conv.}$ of the NN-based functional for Al as it did not converge.}
\label{fig:convergence_3mat}
\end{figure*}

\section{Results and Discussion}
Table \ref{table:benchmark_molecules} presents the benchmark results for molecules. We compared the performance of the XC functional constructed in this work, denoted as “pcNN-based”, with the performance of the XC functional constructed using a simple NN without satisfying any asymptotic constraints, denoted as “NN-based” \cite{nagai2020completing}. Additionally, PBE and SCAN, which are widely used analytical XC functionals in practical calculations, were also compared. 
While the simple NN and analytical XC functionals demonstrated comparable accuracy, pcNN outperformed them even though the training dataset was almost the same as that of NN. 

Table \ref{table:benchmark_solids} presents the benchmark results for solids, which indicate that  pcNN outperformed the other XC functionals. It is remarkable that improvements for solids were achieved using only the molecular training dataset. For both molecules and solids, the mean errors of ML-based functionals (NN, pcNN) were smaller than those of the analytical ones (PBE, SCAN). This indicates that the error of analytical functionals tend to have biased errors, whereas their ML-based counterparts yield relatively random errors even if physical constraints are imposed. 

Figure \ref{fig:convergence_all} presents the number of iterations until the self-consistent KS cycle converged. With the NN-based functional, the self-consistent calculation was not stable and did not converge for some solids. Calculation with the pcNN-based functional, in contrast, was as stable as that with analytical XC functionals, even though the pseudopotentials used in the calculation were adjusted for the convergence of PBE (this was also advantageous for SCAN because its structure resembles PBE). Figure \ref{fig:convergence_3mat} illustrates the process of total-energy convergence in the self-consistent KS cycles of Si (all XC functionals converged), Fe (NN barely converged with 30 iterations), and Al (NN did not converge). When the KS calculation successfully converged, the energy difference decreased exponentially. When NN failed to converge, the total energy oscillated at certain magnitudes. The convergence of pcNN-based seems much more stable than that of NN-based, and resembles SCAN, as its structure directly included SCAN. These results suggest that if a ML functional is trained as a factor to another functional, the convergence behavior will be similar. 

We argue that the improvement and stability of the pcNN-based functional were due to both the training dataset and the physical constraints. As the training dataset, we used the electronic structure of small molecules whose electrons were spatially localized. In addition, some of the asymptotic constraints imposed by the $\Theta$ operation  were derived from metallic solids where electrons were delocalized (e.g., constraints X3 and C1). Thus, the pcNN-based functional was effective for both localized and delocalized features of electrons. In the construction of the NN-based functional, the properties of the delocalized electrons were not referenced. We consider this to be the cause of the instability in the calculation of solids.

\section{Conclusions}
In this study, we proposed a method to analytically impose asymptotic constraints on ML models, and applied it to construct the XC functional of DFT. Our XC functional based on the proposed NN model satisfied physical asymptotic constraints. As a result, the self-consistent solution of the KS equation converged stably and exhibited higher accuracy than existing standard XC functionals, even for solid systems that were not included in the training dataset. This improvement can be attributed to both the
flexibility of the ML model and the regularization by imposing physical constraints. By incorporating these advantages, the performance of the constructed XC functinoal was improved. 

XC functionals that have been widely used in application studies, such as PBE and SCAN, are stable and accurate for a variety of materials. One of their common features is that they satisfy many analytic physical constraints. Similarly, pcNN is designed to satisfy as many physical constraints as possible analytically. Its high stability and accuracy are expected to be useful in practical research in the future.

The ML modeling method presented in this paper is not limited to the construction of XC functionals, but can will also be used for other ML applications where theoretical asymptotic constraints are known.

\begin{acknowledgments}
This work was supported by KAKENHI Grant No.  JP20J20845 from JSPS. Part of the calculation was performed at the Supercomputer System B and C at the Institute for Solid State Physics, the University of Tokyo.

\end{acknowledgments}

\appendix

\section{Activation Functions}
Since the variable range of the output value $F$ is $[0, \infty)$, that of the activation function should also be semi-infinite.
Additionally, it is desirable for the function to be infinitely differentiable because its derivative is substituted in the KS equation (see Eq. \ref{eq:Vxc}).
Considering these requirements, we adopted the softplus \cite{softplus} as the activation function $\phi({\bf x})$ and changed its coefficients as follows to satisfy $\phi(0)=1$ and $\phi'(0)=1$:
\begin{equation}
    \phi({\bf x})=\frac{1}{\log 2}\log\left(1+\exp\left(2{\bf x}\log 2 \right)\right).
    \label{eq:softplus}
\end{equation}
With these adjustments, when all NN weights are set to $0$, $F_{\rm X, C}^{\rm NN}$ converges to $1$, and the total XC functional thus converges to SCAN.

Since $F_{\rm X}$ and $F_{\rm C}$ should be non-negative (see constraints X4 and C5 in Table \ref{table:applied_conditons}), these activation functions were also applied after each $\Theta$ operation (see Figure \ref{fig:total_architecture}):
\begin{equation}
    F=\phi(F_{\rm X, C}^{\rm tot}-1).
    \label{eq:final_processing}
\end{equation}

\section{\label{app:training}Training Procedures}
The loss function in Eq. \ref{eq:lossfunc} is computed by solving the KS equation (Eq.\ref{eq:KS}) with substituting the NN-based XC functional during training. The KS equation is a self-consistent eigenvalue equation, and thus it is technically difficult to calculate its derivative. In this study, to avoid using  gradients, we performed simulated-annealing-based optimization \cite{simulated_annealing,  simulated_annealing2}. We denote all the weights of the matrices and vectors in NN as ${\bf w}$. ${\bf w}$ was randomly perturbated in every trial step. If NN improved the loss value, the perturbation was accepted; otherwise, it was rejected. We set the step size of the random number generation to $\sigma=0.003$ and the imaginary temperature to $T=0.1\Delta_i^{t-1}$. 
This training was performed with 2,560 parallel threads and approximately 300 trial steps for updating per thread on a supercomputer system with AMD EPYC 7702 and 2 GB RAM per core. The overall training procedure in each parallel thread is described in Algorithm \ref{alg:training}. After this iteration, we collected the trained weights ${\bf w}_i^t$ and loss values $\Delta_i^t$ from all the parallel threads and trial steps and sorted them by $\Delta_i^t$. For the top 100 weights, we computed the mean absolute error for the dataset consisting of the AE of 144 molecules. The molecules were small to medium-sized (the largest molecule was benzene), whose accurate WFT  data are available \cite{G2H0}. Finally, the weights with the highest accuracy were adopted. The computational cost of this training was high because the KS equation for each material was solved in each iteration. Recently, Li {\it et al.} \cite{li2021regularizer}, Kasim {\it et al.}\cite{kasim2021learning}, and Dick {\it et al.}\cite{dick2021using}  implemented back-propagation of the KS equation for some systems to perform gradient-based optimization. Although we avoided using these methods owing to the lack of memory in our environment, they have the potential to improve the training efficiency.

\begin{algorithm}[H]
\caption{Pipeline of training the NN in a parallel thread $i$.}
\textbf{Parameter}: Step size $\sigma$, imaginary temperature $T$, and number of iterations $N_{\rm iter.}$.
\begin{algorithmic}[1] 
\STATE Let ${\bf w}_i^0 = {\bf 0}$ and $\Delta_i^0=\infty$.
\STATE Generate random-weight-perturbation $\delta {\bf w}_i^0$ from $\mathcal{N}(0, \sigma)$, where $\mathcal{N}(a, b)$ represents the normal distribution with mean $a$ and standard deviation $b$.
\FOR {$t = 1$ to $N_{\rm iter.}$}
\STATE Substitute the NN-based XC functional with parameter ${\bf w}_i^{t-1}+\delta {\bf w}_i^{t-1}$ into the KS equation and solve it for the energies and densities to calculate the loss value $\Delta_i^t$ by Eq.\ref{eq:lossfunc}.
\STATE Compute $P=\exp(-(\Delta_i^t-\Delta_i^{t-1})/T)$.
\STATE Generate random number $p_i^t\in(0,1)$ from a uniform distribution.
\IF {$P>1$} 
\STATE Set ${\bf w}_i^t={\bf w}_i^{t-1}+\delta {\bf w}_i^{t-1}$ and $\delta {\bf w}_i^{t}=\delta {\bf w}_i^{t-1}$.
\ELSIF {$p<P<1$}
\STATE  Set ${\bf w}_i^t={\bf w}_i^{t-1}+\delta {\bf w}_i^{t-1}$ and generate the random weight perturbation $\delta {\bf w}_i^t$ from $\mathcal{N}(0, \sigma)$.
\ELSE
\STATE Set ${\bf w}_i^t={\bf w}_i^{t-1}$ and $\Delta_i^t=\Delta_i^{t-1}$, and generate the random weight perturbation $\delta {\bf w}_i^t$ from $\mathcal{N}(0, \sigma)$.
\ENDIF
\ENDFOR
\end{algorithmic}
\label{alg:training}
\end{algorithm}

\end{document}